\documentclass[preprint,showpacs,floatfix]{revtex4}
\usepackage{natbib}
\usepackage{graphicx,subfigure,float,graphics}
\usepackage{epsfig}
\usepackage{subfig}

\begin{document}

\title{Proton-neutron pairing and binding energies of nuclei close to N=Z line}

\author{D. Negrea and N. Sandulescu\footnote{corresponding author, email: sandulescu@theory.nipne.ro}}
\affiliation{National Institute of Physics and Nuclear Engineering, 077125 M\v{a}gurele, Romania}
\author{D. Gambacurta}
\affiliation{INFN-LNS, Laboratori Nazionali del Sud, 95123 Catania, Italy }

\begin{abstract}
We  analyse the contribution of isovector and isoscalar  proton-neutron pairing 
to the  binding energies of even-even nuclei with $N-Z=0,2,4$ and atomic mass $20 < A <100$. 
The binding energies are calculated in the mean-field approach by coupling a  Skyrme-type functional 
to an isovector-isoscalar pairing force of zero range. The latter is treated in the framework of quartet 
condensation model (QCM), which conserves exactly the particle number and the isospin. The interdependence 
of pairing and  deformation is taken into account by performing self-consistent Skyrme-HF+QCM calculations 
in the intrinsic system. It is shown  that the  binding energies are not changing much when the isoscalar
pairing is switched on. This fact is related  to the off-diagonal matrix elements of the pairing force, 
which is less attractive for the isoscalar force, and to the competition between the isoscalar and isovector
pairing channels.   
   
\end{abstract}

\maketitle

\section{\label{sec:intro}Introduction}

  In nuclei close to $N=Z$ line, it is usually considered to be important two types of 
  proton-neutron (pn) pairing correlations, corresponding to spin-singlet isovector 
  (S=0,T=1) and spin-triplet isoscalar (S=1,T=0) pn pairs. 
  Due to the isospin invariance of  nuclear forces, the isovector pn pairing is supposed to play
  a similar role as the standard neutron-neutron and proton-proton pairing.  Much less it is
  known however about the  role played by the isoscalar pn pairing in nuclei. In fact, for quite many
  years, a lot of efforts have been focused on finding the fingerprints of isoscalar pn pairing
  correlations in various nuclear observables such as binding energies, high-spin excitations, 
  proton-neutron transfer cross sections, etc. (e.g., see the recent reviews \cite{frauendorf,sagawa}).

  The majority of theoretical studies on pn pairing have been done in the Hartree-Fock-Bogoliubov 
  (HFB) approach. In HFB the pn pairing, both isovector and isoscalar, is  treated together
  with the  like-particle pairing through the generalized
  Bogoliubov transformation (e.g., see \cite{goodman_review,goodman2001,gezerlis}  and the references
  quoted therein). For most of nuclei, the HFB calculations predict T=1  pairing correlations in the ground state.
  The T=0 pairing and the coexistence between T=1 and T=0 pairing is predicted for a few nuclei, 
  but these predictions depend strongly on the chosen parameters and the calculation scheme. 
  It is also  not clear how these predictions are affected by the non-conservation of particle number, 
  the isospin and the angular momentum, which are specific to the HFB calculations done with the
  isovector-isoscalar pairing interactions. To conserve all these quantities in HFB
  calculations is a difficult task and some results on this line exist only for the trivial 
  case of degenerate levels \cite{dobes,romero}.  Realistic beyond-HFB   
  calculations with particle number and angular momentum projections have been done recently, 
  but with the projection performed after the variation \cite{hfb_augusto}. Another source of 
  uncertainty comes from the fact that in the majority of HFB calculations the mean field is
  kept fixed, so it is not taken into account dynamically the competition between  pairing
  and deformation \cite{goodman2001}. This is also the case of the most recent HFB calculations, 
  done on the top of a fixed spherically-symmetric mean field, in which the effect of the deformation
  on pairing is neglected completely \cite{gezerlis} .

  An alternative approach to take into account the isovector-isoscalar pairing correlations
  in mean-field approximations was proposed in Refs \cite{qcm_t0t1_nez,qcm_t0t1_ngz}. 
  In this approach, called Quartet Condensation Model (QCM),  the ground state of N=Z nuclei
  is described as a product of quartets  built by two protons and two neutrons coupled to
  the total isospin T=0. By construction, in QCM the ground state is conserving exactly
   both the particle number and the isospin. When the quartets are built
  with spherically symmetric single-particle states, the QCM ground state has also a 
  well-defined angular momentum \cite{qm_qcm_t0t1}. 

   Previous studies have shown that the QCM approach provides accurate results for
   isovector-isoscalar pairing Hamiltonians which can be solved exactly \cite{qcm_t0t1_nez,
   qcm_t0t1_ngz,qm_qcm_t0t1}.  The purpose of this work is to extend these studies
   to self-consistent mean-field plus pairing calculations and to analyse, within the QCM framework, 
  the contribution of T=1 and T=0 pairing   correlations to the ground state energy of nuclei close to N=Z line. 
  The novel feature of the present calculations  is that they take into account dynamically the
   competition between pairing and deformation in a formalism which conserves exactly both the
  particle number and the isospin.

\section{\label{sec:model} The Formalism}

To calculate the binding energies, we use a self-consistent mean-field plus
pairing formalism. The calculations are done in the intrinsic system defined by an axially
deformed mean-field generated by a Skyrme functional. The pairing correlations
are induced by an isovector-isoscalar pairing force which scatters pairs of nucleons in
time-reversed  states. To evaluate the contribution of pairing correlations to the
binding energies, we employ the  QCM approach introduced in Refs \cite{qcm_t0t1_nez,qcm_t0t1_ngz}. 
For the sake of completeness, the QCM formalism is shortly presented below. 

 The isovector and isoscalar pairing correlations are calculated for a set of axially-deformed 
 single-particle states. They are described by the Hamiltonian \cite{qcm_t0t1_nez},

\begin{eqnarray}
H&=&\sum_{i,\tau=\pm1/2}\varepsilon_{i\tau}N_{i\tau}+\sum_{i,j}V^{(T=1)}_{i,j}\sum_{t=-1,0,1}P_{i,t}^{\dag}P_{j,t}\nonumber\\
&&+\sum_{i,j}V_{i,j}^{(T=0)}D_{i,0}^{\dag}D_{j,0},
\label{ham}
\end{eqnarray}
where $\varepsilon_{i,\tau}$ are the single-particle energies of neutrons ($\tau$=1/2) and protons ($\tau$=-1/2),
while $N_{i,\tau}$ are the particle number operators. The second term is
the isovector pairing interaction expressed by the isovector pair operators
$P_{i,1}^{\dag}=\nu_i^{\dag}\nu_{\bar{i}}^{\dag}$, $P_{i,-1}^{\dag}=\pi_i^{\dag}\pi_{\bar{i}}^{\dag}$, $P_{i,0}^{\dag}=(\nu_i^{\dag}\pi_{\bar{i}}^{\dag}+\pi_i^{\dag}\nu_{\bar{i}}^{\dag})/\sqrt{2}$. The third term is the
isoscalar pairing interaction and $D_{i,0}^{\dag}=(\nu_i^{\dag}\pi_{\bar{i}}^{\dag}-\pi_i^{\dag}\nu_{\bar{i}}^{\dag})/\sqrt{2}$ is the isoscalar pair operator. By $\nu^{\dag}_i$ and $\pi^{\dag}_i$ are denoted the
creation operators for neutrons and protons in the state $i$, while $\bar{i}$ is the time
conjugate of the state $i$. The states $i$, which correspond to the axially deformed mean-field,
are characterised by the quantum numbers $i \equiv \{a_i, \Omega_i \}$, where $\Omega_i$ is the projection of the angular momentum on the symmetry axis. 

By construction, in Eq. (1) the pairs operators have $J_z=0$ but not a well-defined
angular momentum $J$. In fact, when expressed in the laboratory frame, the isovector and the isoscalar intrinsic
pairs can be written as a superposition of pairs with $J={0,2,4, ..}$ and, respectively, $J={1,3,5,..}$. Therefore,
the Hamiltonian (1) takes into account, in an effective way, pairing correlations which are not restricted
only to the standard (J=0,T=1) and (J=1,T=0) channels.

In order to find the ground state energy of the Hamiltonian (1), we employ the quartet condensation
model (QCM). Thus, according to QCM, the ground state of Hamiltonian (1) for even-even $N=Z$ systems
is approximated by the trial state \cite{qcm_t0t1_nez}
\begin{equation}
|QCM\rangle=(A^{\dag}+\Delta_0^{\dag 2})^{n_q}|0\rangle,
\label{qcm}
\end{equation}
where $n_q=(N+Z)/2$, while $|0\rangle$ is the "vacuum" state represented by the nucleons which are 
supposed to be not affected by the pairing interaction. 
The operator $A^{\dag}$ is the isovector quartet built 
by two isovector non-collective pairs coupled to the total isospin $T=0$, i.e.,
\begin{equation}
A^{\dag}=\sum_{i,j}x_{ij}[P^{\dag}_i P^{\dag}_j]^{T=0}.
\end{equation}
Assuming that the mixing coefficients are separable, i.e., $x_{ij}=x_i x_j$, 
the isovector quartet takes the form
\begin{equation}
A^{\dag}=2\Gamma_{1}^{\dag}\Gamma_{-1}^{\dag}-(\Gamma_{0}^{\dag})^{2},
\label{isovq}
\end{equation}
where 
\begin{equation}
\Gamma_{t}^{\dag}=\sum_{i}x_{i}P_{i,t}^{\dag}
\end{equation}
are collective pair operators for neutron-neutron
pairs ($t=1$), proton-proton pairs ($t=-1$) and proton-neutron pairs ($t=0$). 
The isoscalar degrees of freedom are described by the
collective isoscalar pair
\begin{equation}
\Delta_{0}^{\dag}=\sum_{i}y_{i}D_{i,0}^{\dag}.
\label{isosq}
\end{equation}

For even-even systems with $N > Z$ (the case $N<Z$ is treated in the same manner)
the ground state is described by \cite{qcm_t0t1_ngz}
\begin{equation}
|QCM\rangle=(\tilde{\Gamma}_1^{\dag})^{n_N} (A^{\dag}+\Delta_0^{\dag 2})^{n_q}|0\rangle,
\label{totalwf}
\end{equation}
where $n_N=(N-Z)/2$ gives the number of neutron pairs in excess, while $n_{q}=(N+Z-2n_N)/4$ 
denotes the maximum number of quartets which can be formed with $Z$ protons. As in the case of
$N=Z$ nuclei, here by $Z$ and $N$ are denoted the numbers of protons and neutrons above
the $N=Z$ core $|0 \rangle$, which are affected by the pairing interaction. The extra neutrons
are represented by the collective neutron pair
\begin{equation}
\tilde{\Gamma}_{1}^{\dag}=\sum_{i}z_{i}P_{i,1}^{\dag}.
\end{equation}
As can be seen, the structure of the extra pairs, expressed by the mixing amplitudes,
is different from the structure of the neutron pairs which enter in the definition
of the isovector quartet (4).

The QCM states  depend on the mixing amplitudes of the collective pair operators.
They are determined variationally by minimizing the average of the Hamiltonian under the
normalization condition imposed to the trial state. Details about these calculations are 
presented in Ref. \cite{qcm_t0t1_nez} and in the Appendix of Ref. \cite{qcm_t0t1_ngz}.
 
The QCM calculations for the Hamiltonian (1) are performed iteratively with the Skyrme-HF calculations
in a similar way as in the axially-deformed Skyrme+BCS calculations \cite{vautherin}. Thus, at a given
iteration, the QCM equations are solved for the single-particle states generated by the Skyrme
functional. Then, the occupation probabilities of the single-particle states provided by QCM are employed
to get new densities and a new Skyrme functional which, in turn, is generating new single-particle states. 
At the convergence, the binding energy is obtained by adding to the mean-field energy 
the contribution of the pairing energy. The latter is calculated as the average of the pairing force
from which it is extracted out the contribution of self-energy terms. For the like-particle pairing
these terms are
\begin{equation}
E^{mf}_{n(p)}=\sum_i V^{T=1}(i,i) v^4_{i,n(p)},
\end{equation}
while for the pn pairing the expresions are
\begin{equation}
E^{mf}_{pn}(T)=\sum_i V^{T}(i,i) v^2_{i,p} v^2_{i,n}.
\end{equation}
In the expressions above, $v^2_{i,n(p)}$ are the occupation probabilities for neutrons (protons) corresponding to the states
included in the pairing calculations. The terms (9,10), which would renormalize the single-particle energies 
generated by the Skyrme functional, are neglected since in the Skyrme-HF+QCM calculations
the pairing force is a residual interaction acting only in the particle-particle channel. 

In the present calculations, for the isovector-isoscalar pairing interaction we employ a zero range
force of the form:
\begin{equation}
V^T (r_1,r_2)=V^T_0\delta(r_1-r_2)\hat{P}^T_{S,S_z}
\end{equation}
where $\hat{P}^T_{S,S_z}$ is the projection operator on the spin of the pairs, namely, $S=0$ for 
the isovector  force and $S=1, S_z=0$ for the isoscalar  force. The matrix elements of the pairing
interaction (11) for the single-particle states provided by the Skyrme functional  are calculated 
as shown in the Appendix of Ref. \cite{danilo_denis}.

To distinguish between various quantities originating from the pairing interaction (11), in what follows we shall
denote by interaction energy the average of the pairing interaction on the QCM state, by pairing energy
the average of the pairing force without the contribution of the terms (9,10),  while the average
of the latter is denoted by self-energy.

\section{Results}

The Skyrme-HF+QCM formalism  presented above is applied to analyse the effect
of T=1 and T=0 pairing on the binding energies of nuclei with the atomic mass $A=N+Z$ 
between 20 and 100. We consider first the even-even nuclei with $N=Z$, for which the pn
pairing correlations are supposed to be the largest, and then the nuclei 
with $N=Z+2$ and $N=Z+4$. 

\subsection{Calculation scheme} 

To set the calculation scheme for the Skyrme-HF+QCM calculations one needs to chose the 
Skyrme functional, the pairing force and the model space for the pairing
calculations. For the mean field we consider the Skyrme functional UNE1 \cite{une1}.
 The Skyrme-HF calculations have been done with the code EV8 \cite{ev8}, in which the
mean field equations are solved in coordinate space. The mean-field is considered to have
axial symmetry, so the neutron and the proton levels are  double degenerate with respect
to the projection of the angular momentum on the symmetry axis. 

The QCM calculations are performed by solving analytically the QCM equations for the average
of the pairing Hamiltonian and for the norm of the QCM wave function. This has been done by employing
the Cadabra algorithm \cite{cadabra}. In order to keep feasible the analytical derivations, in the QCM
state for the $N=Z$ nuclei (Eq. 2) we have used $n_q=3$ while for the QCM state for  $N>Z$ nuclei (Eq. 7)
 we have taken $n_q=2$.

For the isovector-isoscalar pairing interaction  we employ the delta force given in Eq. (11).
Since the force is of zero range, the pairing calculations should be done with a finite number of
single-particle states from the vicinity of Fermi levels. In the present calculations the active
nucleons are allow to scatter, due to the pairing force, in 10 neutron and 10 proton single-particle
states above the core defined by the QCM states (2,7).

What remains to be chosen are the strengths of the pairing forces, i.e., $V^{T=1}_0$ and $V^{T=0}_0$, or,
 equivalently, the strength of the isovector pairing  $V_0=V^{T=1}_0$ and the ratio $w= V^{T=0}_0/V^{T=1}_0$.
 How to fix these parameters is a non trivial task because there are not observables which to be related
 unambigously to isovector or to isoscalar pairing. Moreover, as shown in the previous QCM calculations,
 the T=1 and T=0 pairing correlations always coexist and they are very difficult to disentangle because the
 isovector and the isoscalar counterparts of the QCM states (2,7) have a large overlap
\cite{qcm_t0t1_nez,qcm_t0t1_ngz}. The alternative
 we have chosen here is to perform calculations with various parameters and to keep those
 for which the differences between the calculated and experimental binding energies are the smallest. 
 More precisely, we have first calculated the binding energies of a few representative even-even N=Z
 nuclei with $V_0=\{300,350,400,465\}$ MeV/fm$^3$ and $w$=0. Then, for a given $V_0$, we have turned on
 the isoscalar pairing force by increasing $w$  until the value $w$=2. 

\begin{figure*}[h]
\begin{center}
\includegraphics[width=0.50\textwidth, angle=-90]{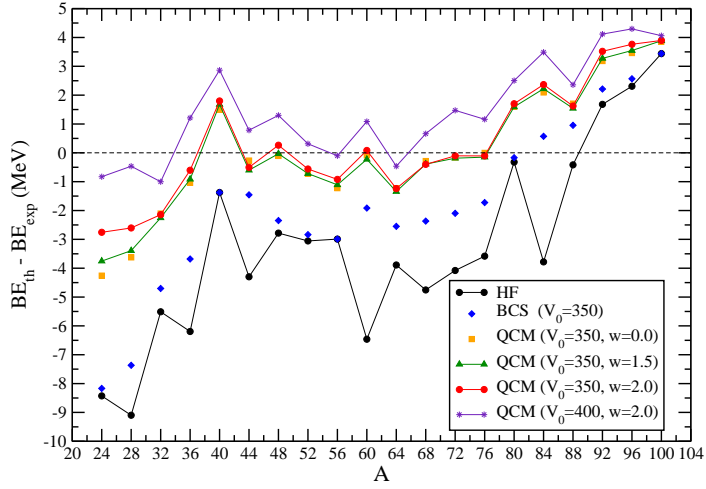}
\end{center}
\caption{Binding energies residuals, in MeV, for even-even N=Z nuclei as a function of A=N+Z.
The results correspond to the pairing forces indicated in the figure.}
\end{figure*}

\renewcommand{\theequation}{3.\arabic{equation}}

\subsection{Pairing and binding energies of $N=Z$ nuclei}

The most representative results for the binding energies are presented in Fig. 1. The figure shows the binding energies residuals,
i.e., the difference between the theoretical and experimental binding energies. The parameters employed in the
calculations are indicated in the figure.  In what follows, we shall focus on the results corresponding to the pairing force of strength $V_0$=350.
First of all, it can be seen that the Skyrme-HF results, obtained by using the equal filling approximation, 
underestimate the 
binding energies by about 3-4 MeV in the middle mass region, while for the nuclei with $A > 90$ the
calculated binding energies are larger than the experimental ones. As expected, the Skyrme-HF+BCS
calculations, which take into account only the neutron-neutron (nn) and proton-proton (pp) pairing, is smoothing
 out the fluctuations of the HF results caused by the shell effects. For the N=Z nuclei  with 
 $60 < A < 80 $, where the HF fluctuations are small, in BCS approximation the residuals are 
 decreasing by about 1 MeV compared to the HF values.

\begin{figure}[h]
\centering
\includegraphics[width=0.35\textwidth]{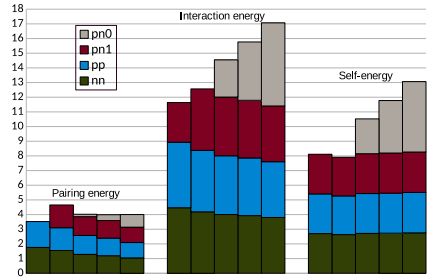}
\caption{Pairing energy, interaction energy and self-energy, in MeV, for $^{64}$Ge. 
 From the left to the right are shown, for each quantity, the PBCS result and the
 QCM results for w=$\{0.0,1.0,1.5,2.0\}$. $pn0$ and $pn1$ indicate the T=0 and T=1 pn channels.}
\end{figure}

 From  Fig. 1 it can be seen that the binding energies are increasing significantly when are taken into
 account the isovector pn  pairing correlations, treated in the QCM approach.  On the other hand, except
 for $A$=24 and $A$=28, the effect of the isoscalar pn  pairing on the binding energies is 
 surprisingly small. This fact is caused  by the competition between various pairing 
 channels and between pairing and mean field. As an example, we discuss in detail the results
 for the nucleus $^{64}$Ge, which is illustrating a typical case.

\begin{figure*}[h]
\centering
\mbox{\subfigure{\includegraphics[width=0.40\textwidth, angle=-90]{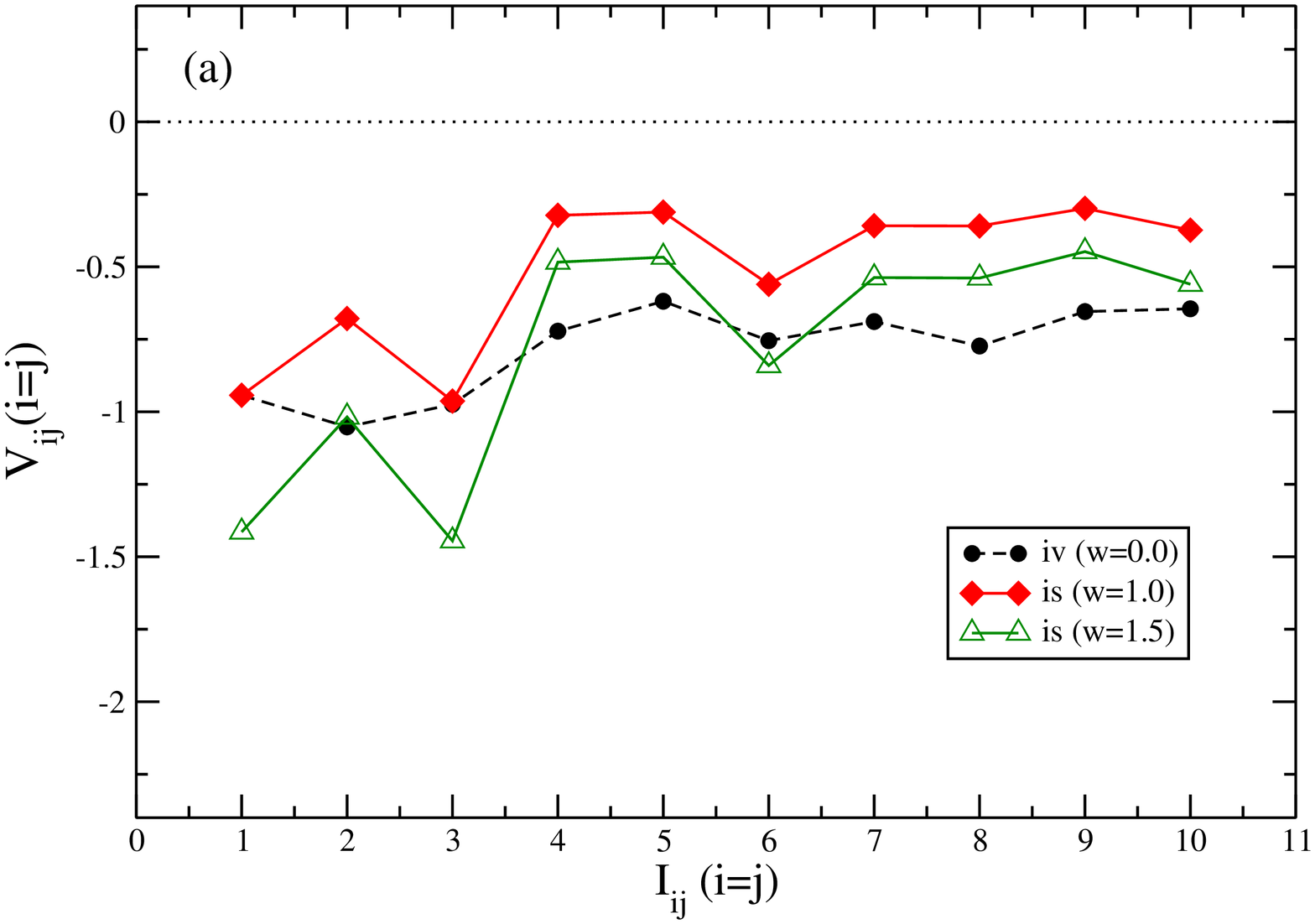}}\quad
\subfigure{\includegraphics[width=0.40\textwidth, angle=-90]{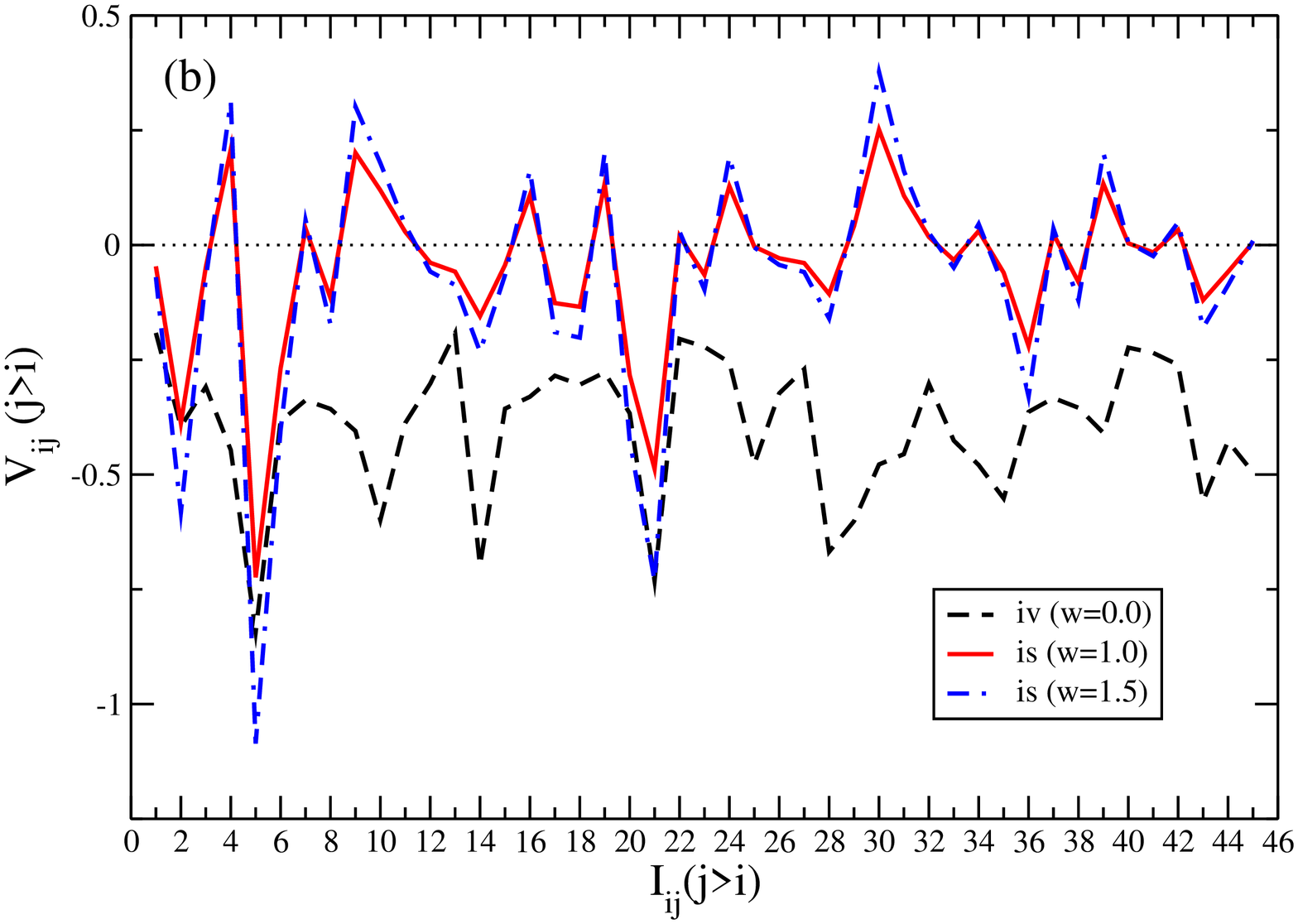}}}
\caption{Diagonal (a) and non-diagonal (b) matrix elements  of the isovector and
 isoscalar  pairing force for  $^{64}$Ge. The quantity $I_{ij}$ 
 enumerates the pair indices of $V_{ij}$.}
\end{figure*}

  In Fig. 2 are shown the pairing energies for $^{64}$Ge provided by 
 the QCM calculations for $V_0$=350 and $w$=$\{0.0,1,1.5,2\}$. In order to disentangle the pairing and the mean
 field effects, the results shown in Fig. 2 correspond to the calculations done on the top of the fixed mean 
 field generated by the Skyrme-HF calculations. As a reference, in the same figure we have included also the
 pairing energies provided by the particle-number  projected-BCS (PBCS) approach in which the variation is done
 after the projection.  The PBCS wave function is taken as a product between a neutron and
 a proton pair condensate, so it does not take into account the isovector pn pairing correlations.
 The latter are taken into account in the isovector QCM approach ($w$=0) and, as expected, they increase
 the total pairing energy compared to PBCS. By contrast, the like-particle pairing
 energies are larger in PBCS than in QCM. This is due the fact that in isovector QCM the 
 like-particle pairing is competing with the isovector pn pairing because they build up correlations
 by sharing the same model space. For the same reason, the like-particle and isovector pn pairing energies
 are decreasing further when the isoscalar pn channel is switched on. 
 Yet, as seen in Fig. 2, in this case the decrease of the isovector pairing is not compensated by the
 pairing energy gained by opening the isoscalar channel. On the other hand, the contribution of the isoscalar
 pn channel to the interaction energy is increasing rapidly with the scaling factor $w$, becoming almost
 equal to the isovector pn  channel for $w=1.5$.  However, as seen from Fig. 2, most of the interaction energy
 in the isoscalar channel is  coming from the self-energy. As a result, the contribution of the isoscalar
 pn pairing to the total pairing, in which it is not included the self-energy, is  reduced significantly,
 much more than  for the isovector pn pairing. This behavior can be traced back to the matrix elements 
 (m.e.) of the pairing interaction,  shown in Fig. 3.  Thus, as seen in Fig. 3a, by increasing the scaling factor, 
 some diagonal m.e. of the isoscalar pn pairing become larger than the isovector ones.
 However, the contribution of diagonal m.e. is drastically reduced when the self-energy terms are subtracted. 
 Therefore, due to the subtraction, the dominant contribution to the pairing energies 
 comes from the off-diagonal m.e., shown in Fig. 3b. It can be noticed that, in average, the m.e. of the isoscalar interaction are 
 smaller than the m.e. of the isovector interaction and, more importantly, some of the isoscalar m.e. 
 are positive. Due to these reasons, the contribution of the isoscalar pairing force to the pairing correlations
 is  not increasing significantly with the scaling factor. In addition, in self-consistent Skyrme-HF+QCM
 calculations, the variation of the pairing energies can be  compensated  by the mean field energy.
 In this case, when the isoscalar channel is turned on, the mean field energy is increasing 
 by about 530 keV for $w$=1.5 and by 540 keV for $w$=2.0, while the total pairing energy is decreasing,   
 relative to the isovector pairing, by about the same quantity. As a consequence, as seen in Fig. 1,
 the total binding energy of $^{64}$Ge is not changing much when the isoscalar pairing force is turned on.  

\begin{figure*}[h]
\centering
\mbox{\subfigure{
\includegraphics[width=0.45\textwidth]{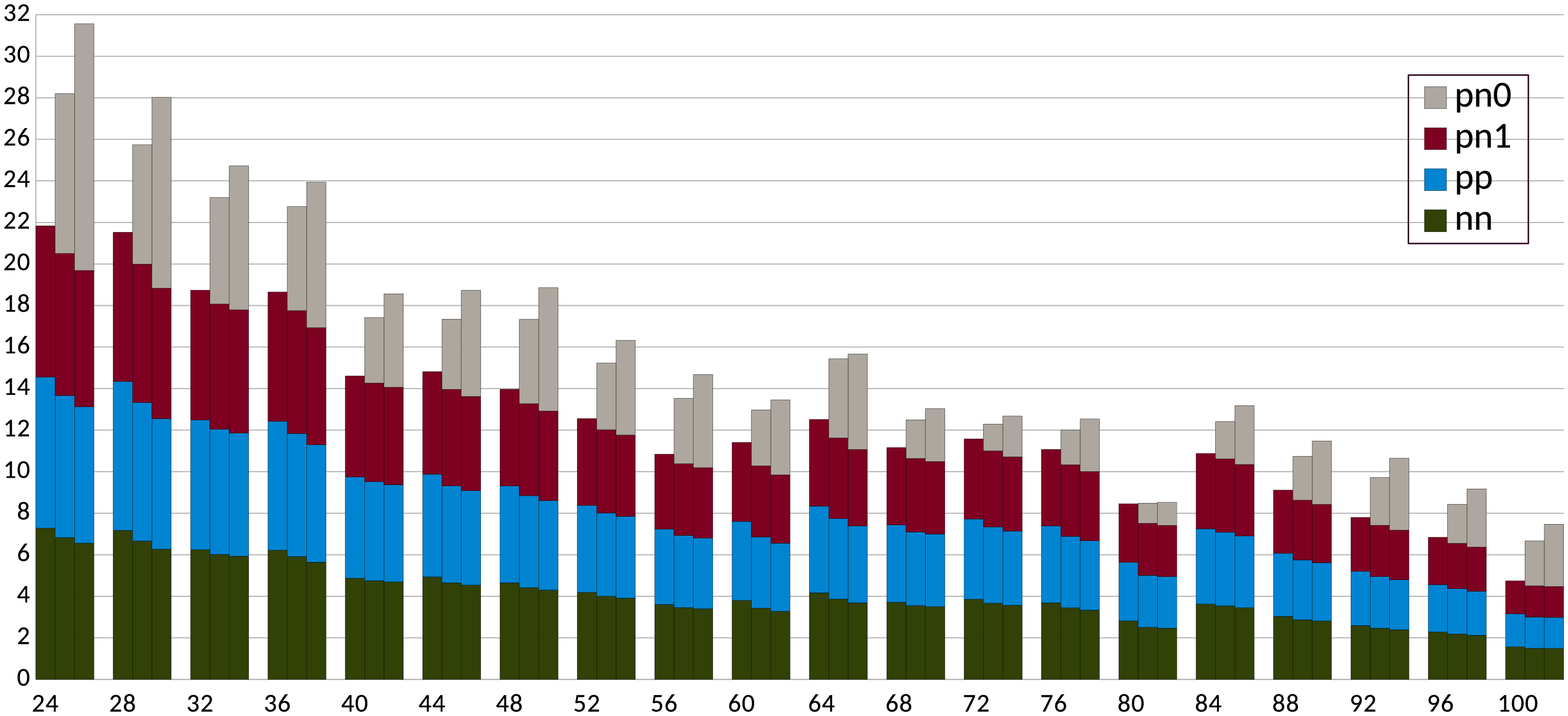}}\quad
\subfigure{
\includegraphics[width=0.45\textwidth]{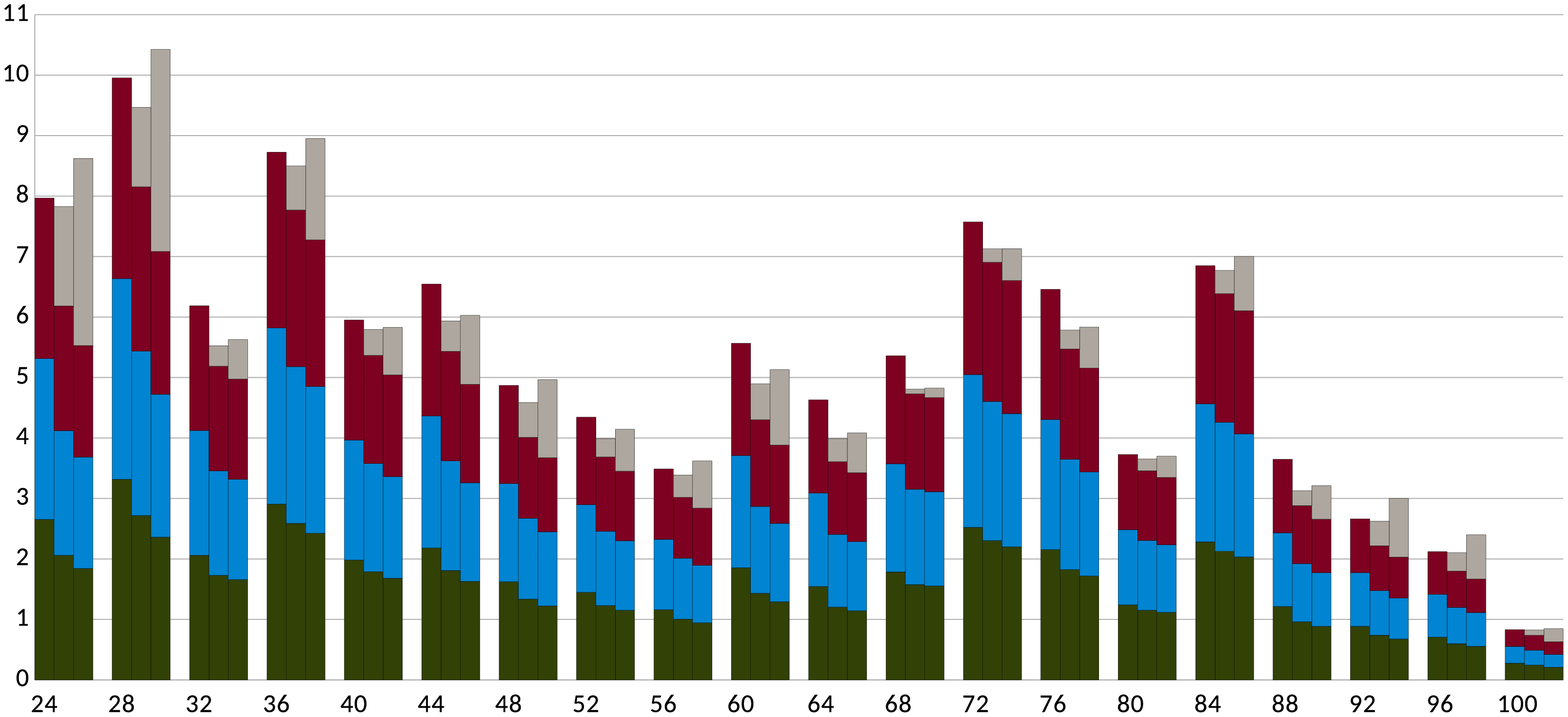}}}
\caption{Interaction energies  (left) and pairing  energies (right), in MeV,  for $N=Z$ nuclei.
  For each nucleus are shown, from the left to the right, the  results for w=$\{0.0,1.5.2.0\}$}
\end{figure*}

The pairing energies and the interaction energies  provided by the self-consistent calculations for all $N=Z$ nuclei
 are shown in Fig. 4. It can be seen that in the majority of nuclei these quantities have a similar
 pattern as in the example discussed above. In particular, we have found that when the isovector and
 the isoscalar interactions have the same strength ($w=1$) the total pairing energy is smaller compared
 to the isovector pairing ($w=0$) in all $N=Z$ nuclei considered in this study. 

 An interesting feature seen in Fig. 4 is that the pairing energies are significant for double magic nuclei $^{40}$Ca, $^{56}$Ni, for which the BCS approximation predicts no pairing. The fact that there are
pairing correlations in $^{40}$Ca was also pointed out in Ref. \cite{volya}. It is worth mentioning as 
well that for $^{40}$Ca and $^{56}$Ni the pairing gaps extracted from the odd-even mass difference 
are large, of 3.6 MeV and 3.2 MeV, respectively. The gaps are also quite large, of the order of  1 MeV, 
in the neighboring odd-even isotopes.

\begin{figure*}[h]
\centering
\mbox{\subfigure{\includegraphics[width=0.40\textwidth, angle=-90]{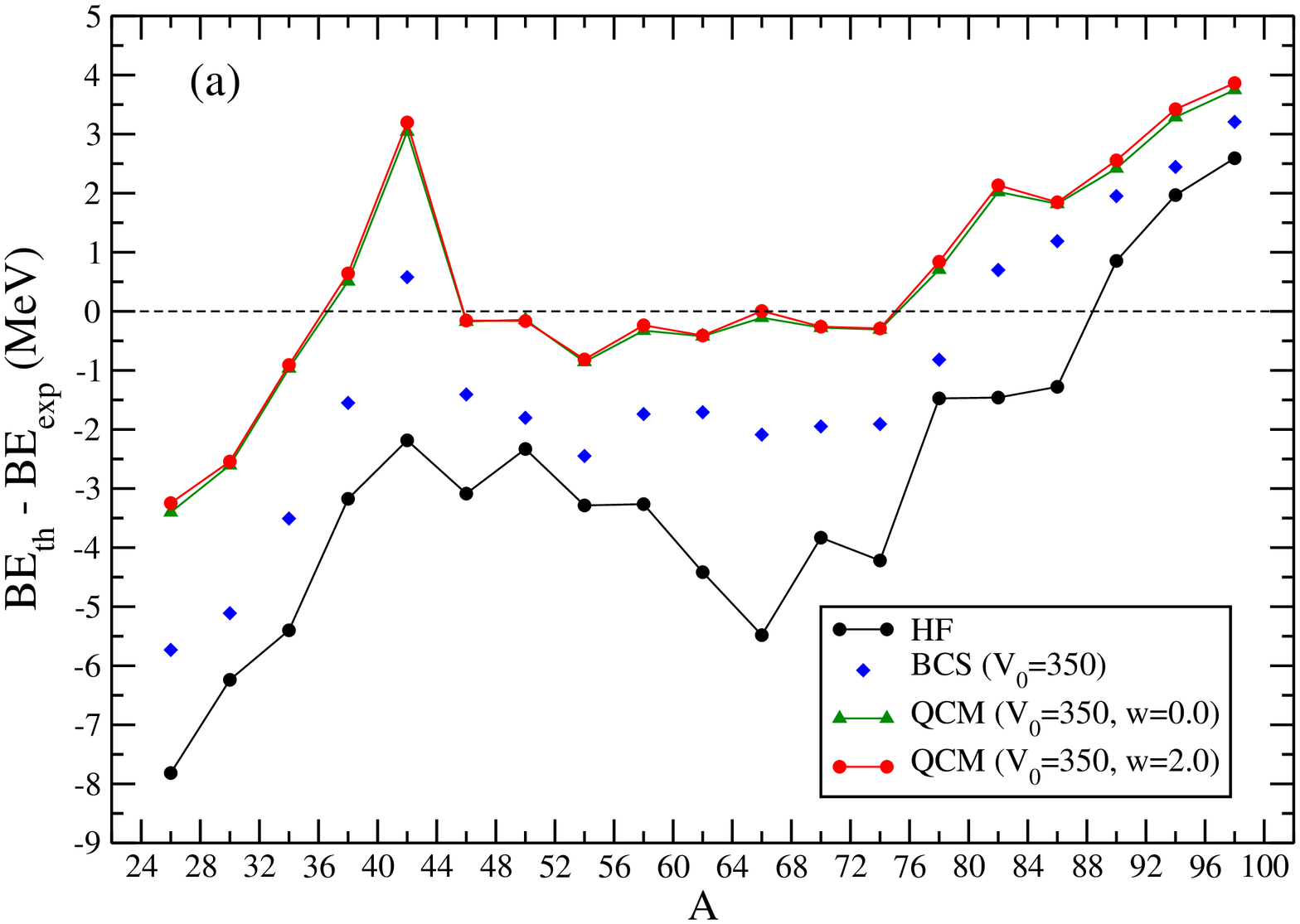}}\quad
\subfigure{\includegraphics[width=0.40\textwidth, angle=-90]{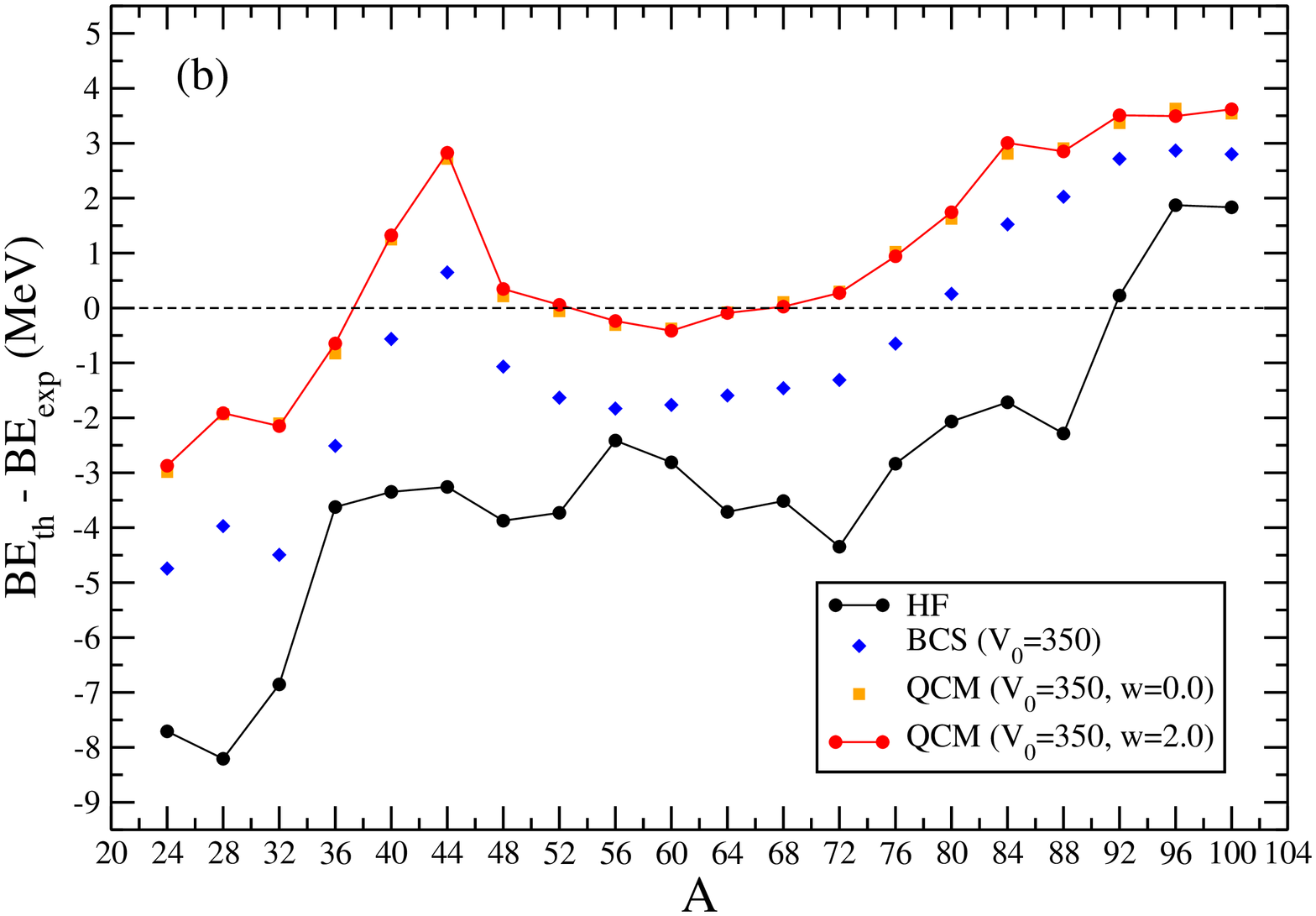}}}
\caption{The residuals, function of $A=N+Z$, for  $N=Z+2$ (a) and $N=Z+4$ (b) nuclei.}
\end{figure*}

\begin{figure*}[h]
\centering
\mbox{\subfigure{
\includegraphics[width=0.45\textwidth]{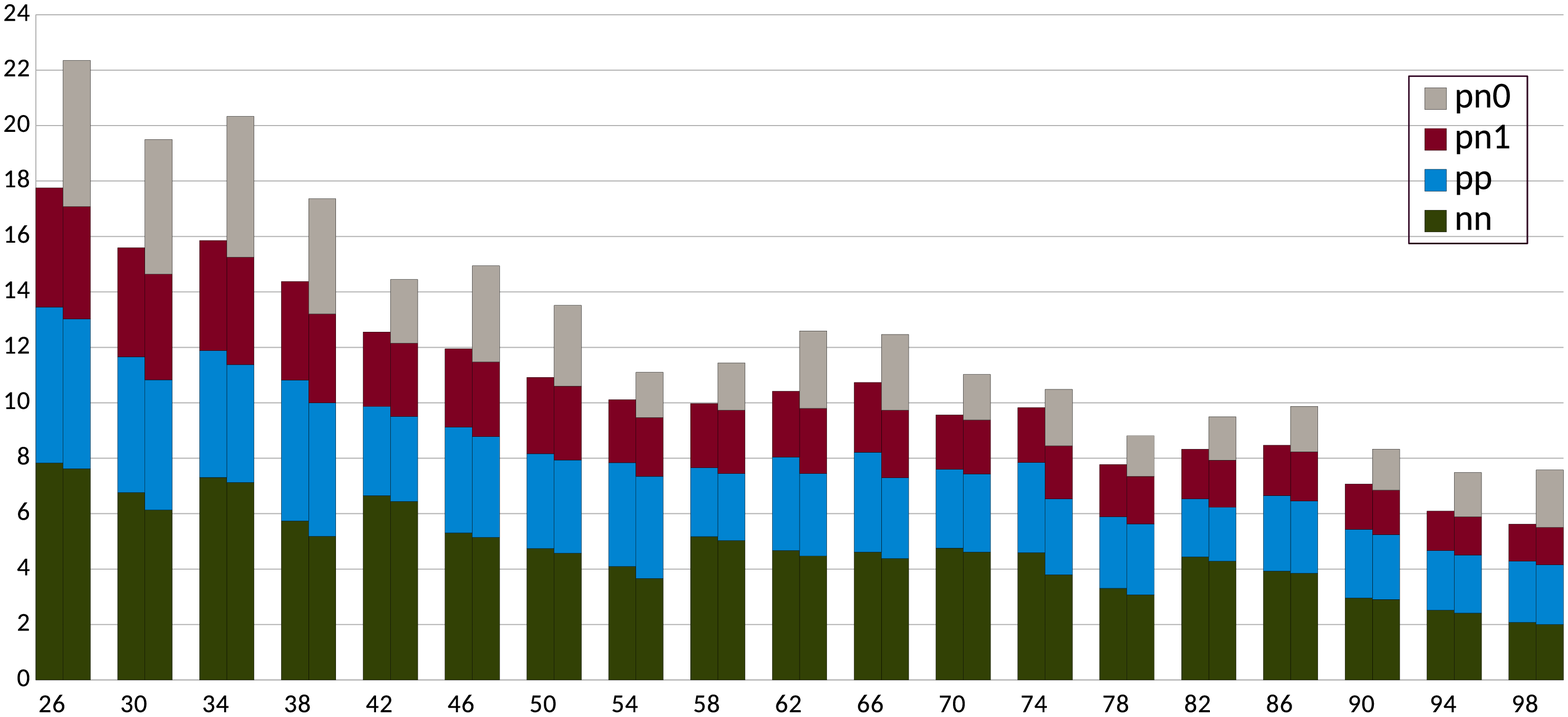}}\quad
\subfigure{
\includegraphics[width=0.45\textwidth]{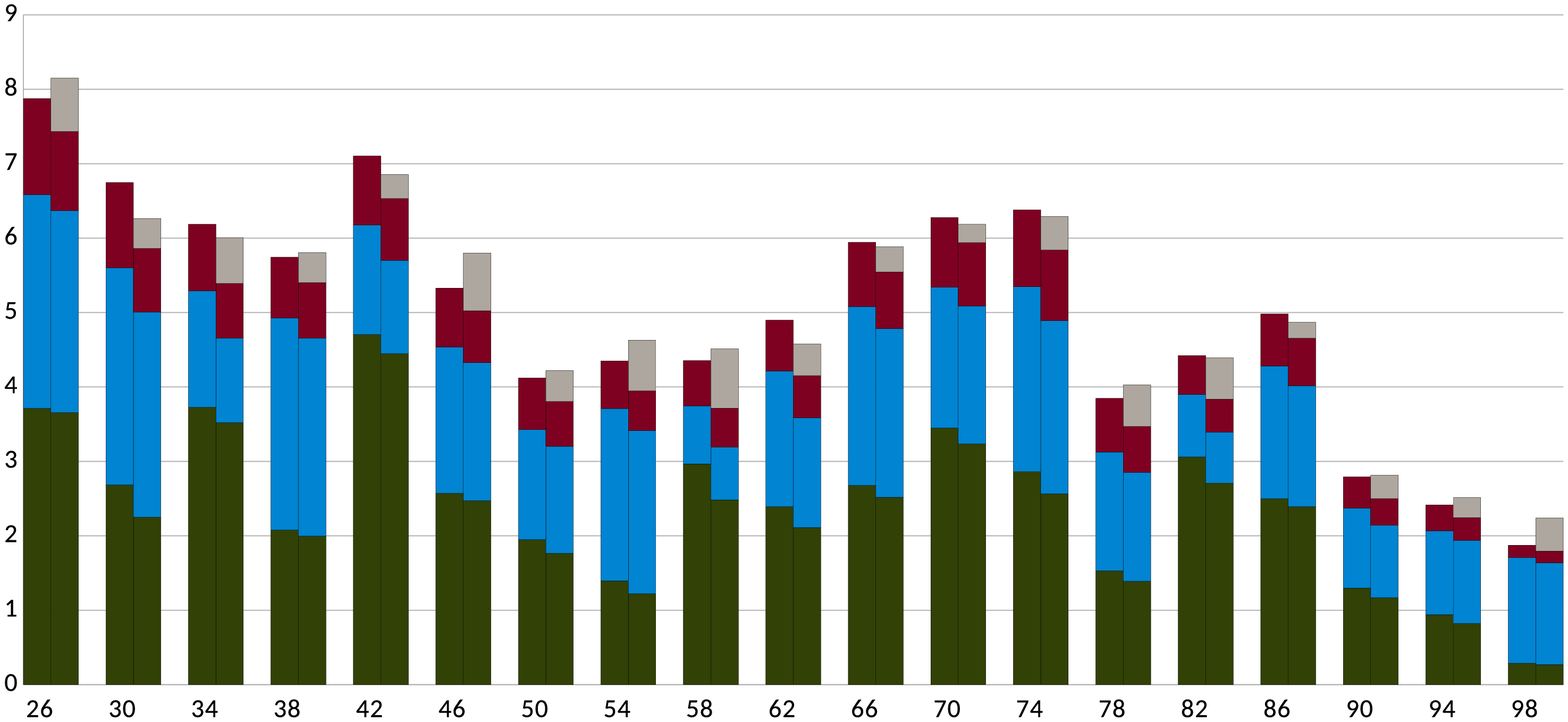}}}
\mbox{\subfigure{
\includegraphics[width=0.45\textwidth]{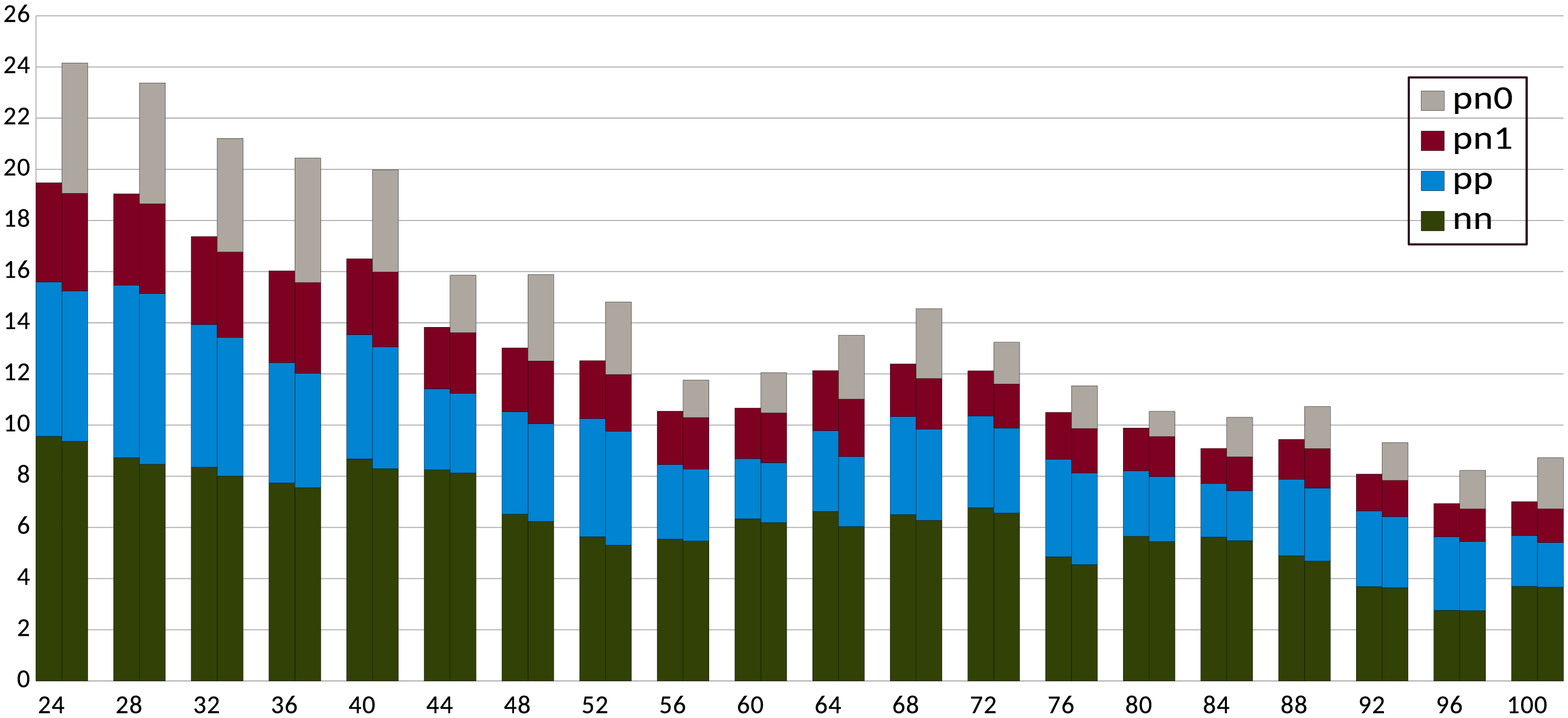}}\quad
\subfigure{
\includegraphics[width=0.45\textwidth]{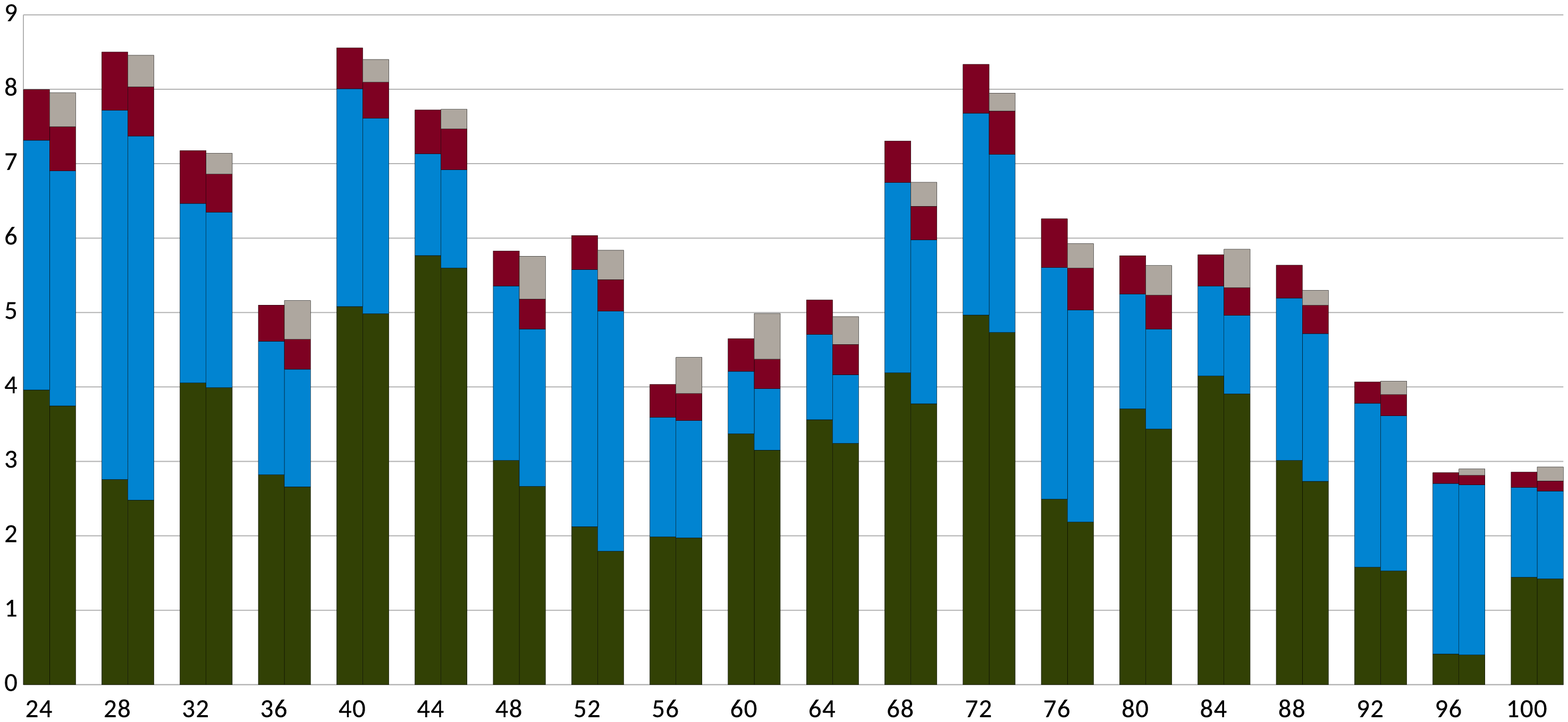}}}
\caption{Interaction energies (left) and pairing energies (right), in MeV,
 function of $A=N+Z$, for the nuclei with $N=Z+2$ (top) and $N=Z+4$ (bottom). 
 For each nucleus are shown, from  the left to the right, the results for w= $\{0.0,2.0 \}$ }
\end{figure*}

\begin{figure*}[h]
\centering
\mbox{\subfigure{
\includegraphics[width=0.45\textwidth]{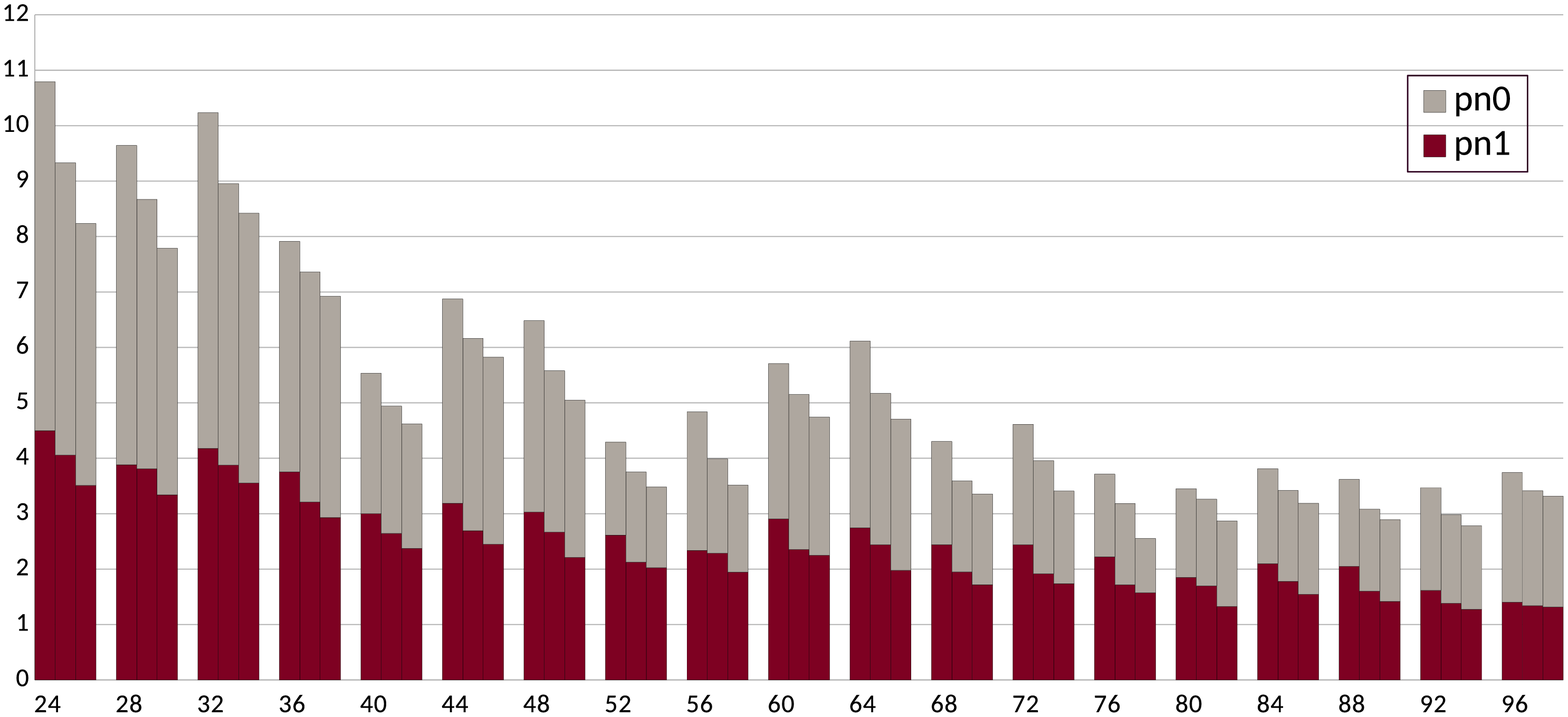}}\quad
\subfigure{
\includegraphics[width=0.45\textwidth]{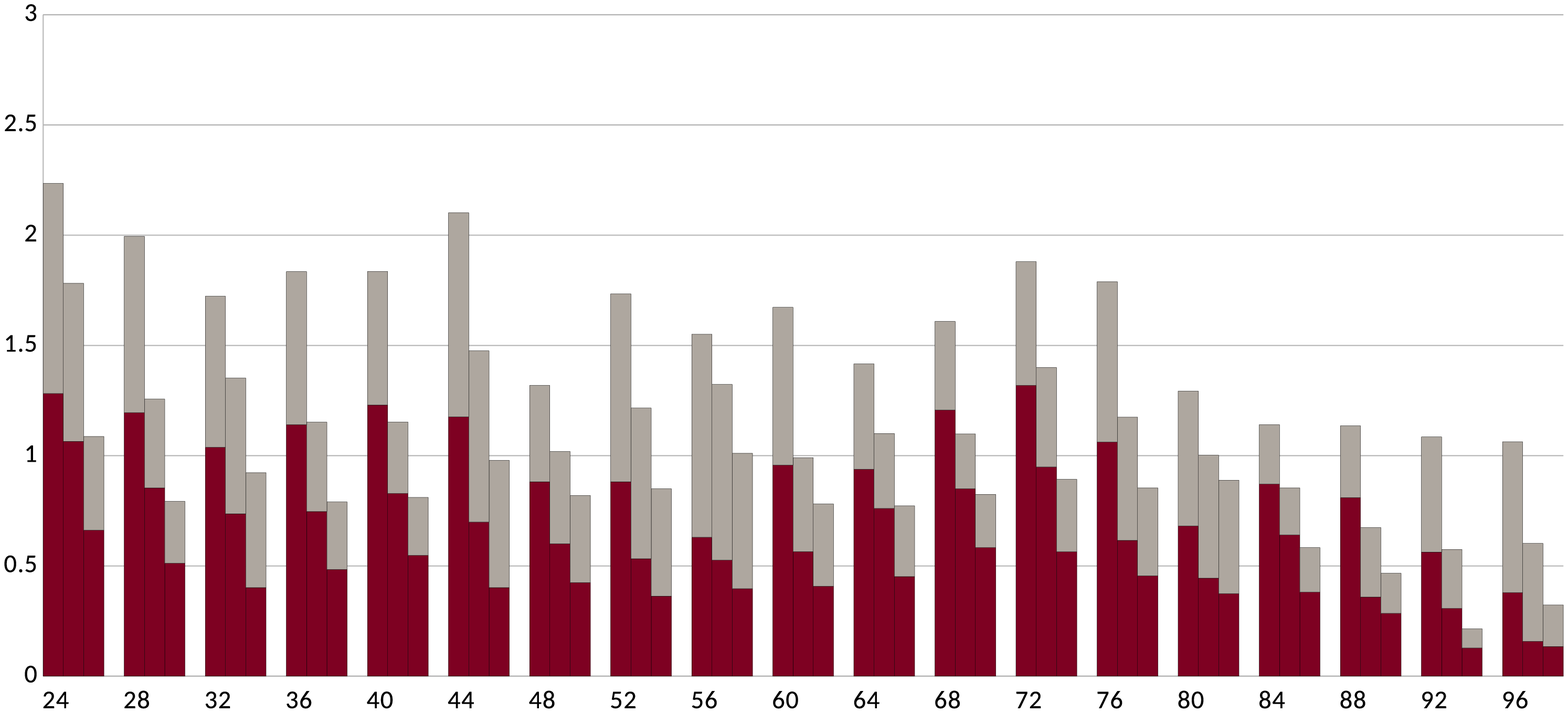}}}
\caption{Interaction energies (left) and pairing energies (right) for T=0 and T=1 pn pairing and for $w=2.0$.
 The results, from the left to the right, are for the nuclei with $N=Z$, $N=Z+2$ and $N=Z+4$. 
 On x-axis is indicated the atomic mass of $N=Z$ nuclei.}
\end{figure*}

\subsection{Pairing and binding energies of $N>Z$ nuclei}

  To study how the pairing correlations are affected by the extra neutrons added to N=Z nuclei, we take 
 as examples the nuclei with $N=Z+2$ and $N=Z+4$ and with atomic mass $ 20 < Z < 100 $ . The binding energy
 residuals  for these nuclei are given in Fig. 5. Are shown the results for the pairing force with $V_0$=350 and $w=\{0.0,2.0\}$. 
 The contribution of the pairing energies to the binding energies is displayed in Fig. 6. For reference,
 in Fig. 6 are given also the interaction energies. The latter are increasing significantly when the
 isoscalar  pairing is turned on. On the contrary, this is not the case for the pairing energies. 
 The reasons for that are the same as in the case of $N=Z$ nuclei: (i) the off-diagonal m.e. of the 
 interaction are less attractive for the isoscalar force; (iii) the pairing channels are competing
 with each other and also with the mean field. As a result, the binding energies of $N>Z$ nuclei shown
 in Fig. 5  change very little when the isoscalar pairing is switched on. This does not mean, 
however, that
 the isoscalar pairing correlations do not contribute to the binding energy of $N>Z$ nuclei. This can be seen 
 clearly from Fig. 7, which shows how the pn pairing energies and interaction energies are
 changing by adding neutrons to the $N=Z$ nuclei. In both T=0 and T=1 channels these
 energies are decreasing when more neutrons are added. However, they are not vanishing, including for
  the nuclei with $N=Z+4$, and they coexist in all the nuclei. In fact, this is happening not 
  only for the large isoscalar strength considered here, but also for any QCM calculations with an 
  isovector-isoscalar pairing force with $w>0$.

\section{Summary and Conclusions}

We have discussed the contribution of isovector and isoscalar pairing on binding energies of
$N=Z$ nuclei and of $N>Z$ nuclei with $N=Z+2$ and $N=Z+4$. The binding energies have been
obtained by performing self-consistent Skyrme-HF+QCM calculations in the intrinsic system.
An interesting aspect pointed out by these calculations is the strong interdependence 
between all types of pairing correlations. In particular, when the isoscalar pn pairing
channel is switched on, the pairing  correlations are redistributed among all the pairing 
channels without changing significantly the total pairing energy. Due to this reason,
for the majority of $N \approx Z$ nuclei the binding energy is not affected much when
the isoscalar pairing channel is switched on. Yet, in all the calculations which include
both the isovector and the isoscalar pairing forces, the isoscalar pairing correlations
contribute significantly to the binding energies and coexist always with the isovector
pn pairing.  This feature, discussed already in the previous studies  \cite{qcm_t0t1_nez,qcm_t0t1_ngz},  
is related to the exact conservation of the particle number and the isospin by the QCM approach. 

The present Skyrme-HF+QCM calculations are based on two approximations 
which should be further checked and improved. Thus, on one hand, since the
calculations are done in the intrinsic system, the ground states have not
a well-defined angular momentum. On the other hand, in the isoscalar pairing channel
are considered only proton-neutron pairs in time-reversed states. This is a rather common
choice when the pairing calculations are done with a deformed mean field
\cite{pacearescu}. In principle, should be also introduced the correlations corresponding
to proton-neutron pairs with $S=1,S_z=\pm 1$. How to treat these correlations in self-consistent
Skyrme-HF+QCM calculations is a non-trivial task which will be addressed in a future study.

\begin{acknowledgments}
This work was supported by a grant of Romanian Ministry of Research and Innovation, CNCS - UEFISCDI,
project number PCE 160/2021, within PNCDI II.
\end{acknowledgments}


\begin{thebibliography}{99}
\bibitem{frauendorf} S. Frauendorf and A. O. Macchiavelli, Prog. Part. Nucl. Phys. {\bf 78}, 24-90 (2014).
\bibitem{sagawa} H. Sagawa, C. L. Bai, and G. Colo, Phys. Scripta {\bf 91}, 083011 (2016). 
\bibitem{goodman_review} A. L. Goodman, Adv. Nucl. Phys. {\bf 11}, 263 (1979).
\bibitem{goodman2001} A. L. Goodman, Phys. Rev. C {\bf 63}, 044325 (2001).
\bibitem{gezerlis} A. Gezerlis, G.-F. Bertsch, Phys. Rev. Lett. {\bf 106}, 252502 (2011).
\bibitem{dobes} J. Dobes, and S. Pittel, Phys. Rev. C {\bf 57}, 688 (1998).
\bibitem{romero}  A. M. Romero, J. Dobaczewski, A. Pastore, Phys. Lett. B {\bf 795}, 177 (2019).
\bibitem{hfb_augusto} E. Rrapaj, A. O. Macchiavelli, A. Gezerlis, Phys. Rev. C {\bf 99}, 014321 (2019).
\bibitem{qcm_t0t1_nez} N. Sandulescu, D. Negrea, D. Gambacurta, Phys. Lett. B {\bf 751}, 348 (2015).
\bibitem{qcm_t0t1_ngz} N. Negrea, P. Buganu, D. Gambacurta, N. Sandulescu, Phys. Rev. C {\bf 98}, 064319 (2018).
\bibitem{qm_qcm_t0t1} M. Sambataro and N. Sandulescu, Phys. Rev. C {\bf 93}, 054320 (2016).
\bibitem{vautherin} D. Vautherin, Phys. Rev. C {\bf 7}, 296 (1973).
\bibitem{danilo_denis} D. Gambacurta, D. Lacroix, Phys. Rev. C {\bf 91}, 014308 (2015).
\bibitem{une1} K. Kortelainen et al,  Phys. Rev.C {\bf 85}, 024304 (2012).
\bibitem{ev8} P. Bonche, H. Flocard, P.-H. Heenen, Comp. Phys. Comm. {\bf 171}, 49 (2005).
\bibitem{cadabra} V. V. Baran, D. S. Delion, Phys. Rev. C{\bf 99}, 03133 (2019); https://cadabra.science.
\bibitem{volya} A. Volya, V. Zelevinsky, Phys. Lett. B {\bf 574}, 27 (2003).
\bibitem{pacearescu} F. Simkovic, C. Moustakidis, L. Pacearescu, A. Faessler, Phys. Rev. C {\bf 68},
 054319 (2003).
\end{thebibliography}
\end{document}